\documentclass[10pt,a4paper]{article}

\usepackage{amsmath,amssymb,graphicx}

\title{\bf The Peculiar Status of the Second Law of Thermodynamics and 
the Quest for its Violation}

\author{Germano D'Abramo\\
{\small SpaceDyS S.r.l., Navacchio di Cascina, Pisa and}\\
{\small IAPS - Istituto Nazionale di Astrofisica, Via Fosso del Cavaliere, 
100, 00133, Roma, Italy}\\
{\small {\tt dabramo@spacedys.com},  {\tt Germano.DAbramo@iaps.inaf.it}}}

\date{{\it Studies in History and Philosophy of Modern Physics}, to appear}

\begin{document}

\maketitle

\begin{abstract}

Even though the second law of thermodynamics holds the supreme position 
among the laws of nature, as stated by many distinguished scientists, 
notably Eddington and Einstein, its position appears to be also quite 
peculiar. Given the atomic nature of matter, whose behaviour is well 
described by statistical physics, the second law could not hold 
unconditionally, but only statistically. It is not an absolute law. As a 
result of this, in the present paper we try to argue that we have not 
yet any truly cogent argument (known fundamental physical laws) to 
exclude its possible macroscopic violation. Even Landauer's 
information-theoretic principle seems to fall short of the initial 
expectations of being the fundamental `physical' reason of all Maxwell's 
demons failure. Here we propose a modified Szilard engine which operates 
without any steps in the process resembling the creation or destruction 
of information. We argue that the information-based exorcisms must be 
wrong, or at the very least superfluous, and that the real physical 
reason why such engines cannot work lies in the ubiquity of thermal 
fluctuations (and friction).

We see in the above peculiar features the main motivation and rationale 
for pursuing exploratory research to challenge the second law, which is 
still ongoing and probably richer than ever. A quite thorough (and 
critical) description of some of these challenges is also given.\\

\end{abstract}

\section{Introduction}

In its classical and phenomenological formulation, the second law of 
thermodynamics states that ``it is impossible to construct a device 
that, operating in a cycle, will produce no effect other than the 
extraction of the heat from a cooler to a warmer body'' (Clausius 
formulation) or, equivalently, that ``it is impossible to construct a 
device that, operating in a cycle, will produce no effect than the 
extraction of heat from a reservoir and the performance of an equivalent 
amount of work'' (Kelvin--Planck formulation).

\begin{figure}[t]
\begin{center}
\includegraphics[width=6cm]{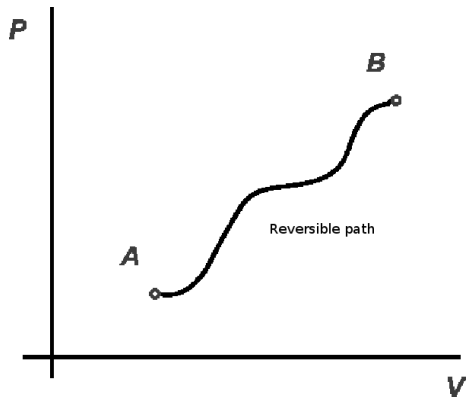}
\end{center}
\caption{Transformation of a thermodynamic system from state $A$ to 
state $B$ through a reversible path in the Volume--Pressure $(V,P)$ 
diagram. The third thermodynamic variable, temperature $T$, depends on 
$V$ and $P$ through the equation of state of the system.}
\label{fig1}
\end{figure}

Then, the thermodynamic state-function entropy $S$ was discovered. It is 
defined up to an additive constant as the following integral through a 
quasi-static reversible path (Fig.~\ref{fig1}),

\begin{equation}
\Delta S=S_B-S_A=\int_A^B\frac{\delta Q}{T},
\label{eq0}
\end{equation}
where $A$ and $B$ are two points in space-state of a thermodynamic 
system, $T$ is the absolute temperature of the system, $\delta Q$ is the 
inexact differential of the heat $Q$ (the heat gained by the system).

The second law can be stated in the well-known {\em increasing entropy} 
formulation: whenever an adiabatically isolated system evolves from 
equilibrium state $A$ to equilibrium state $B$, the variation of entropy 
$\Delta S$ cannot be negative, $\Delta S\geq 0$~\cite{ferm}.

\section{The status of the second law}

Even though the second law ``holds the supreme position among the laws 
of Nature'' (in Eddington's own words~\cite{edd}), its position appears 
to be also quite peculiar. When the theory of statistical physics was 
developed by Maxwell, Boltzmann and others, it became clear very quickly 
that the second law of thermodynamics could not hold unconditionally, 
but only statistically. The Brownian motion is a well-known macroscopic 
example of that, as early noted by Poincar\`e. In other words, entropy 
of isolated systems is not forbidden to decrease, but in all processes 
the {\em probability of continuous and macroscopically significant (and 
also able to provide usable work) entropy decrease is extremely small}.

\begin{figure}[t]
\begin{center}
\includegraphics[width=10cm]{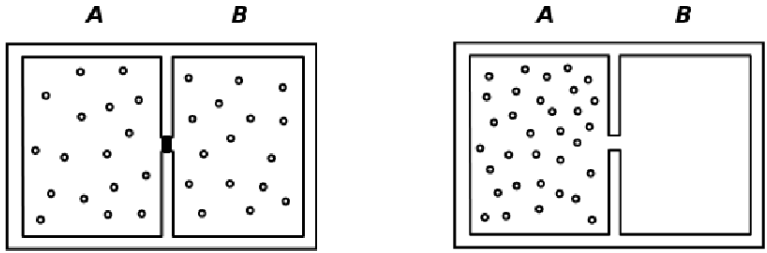}
\end{center}
\caption{The gas-in-two-chambers thought experiment described in the text.}
\label{fig2}
\end{figure}

Consider, for instance, a container separated into two sections, $A$ and 
$B$, by a diaphragm. Both chambers contain the same amount of ideal gas, 
e.g.~$10^{23}$ particles each\footnote{It is well known that 
22.414~liters of gas at standard conditions ($T=263.15\,$K, 
$P=1.0235\times 10^5\,$Pa) contain $6.023\times 10^{23}$ molecules 
(Avogadro's Number). Hence, any macroscopic volume of gas we deal with 
in real life contains no less than $10^{20}\div 10^{23}$ molecules.}, 
and are at the same temperature $T$ (Fig.~\ref{fig2}). If the partition 
which separates chamber $A$ from chamber $B$ is removed, then nothing 
prevents particles in chamber $A$ from freely moving to chamber $B$, and 
vice-versa. Assuming the interaction between particles negligible (we 
are dealing with an ideal gas), the behavior of each particle should be 
uncorrelated with respect to every other particle and there is a 
non-zero chance that, at some moment, all the particles of both chambers 
are confined in chamber $A$. If one observes the system, the probability 
of finding a specific particle in chamber $A$ is obviously equal to 
$\frac{1}{2}$, thus the probability that, at some moment, {\em all} the 
particles are in chamber $A$ is,

\begin{equation}
P_{A\&B\to A}=\biggl(\frac{1}{2}\biggr)^{2\times 10^{23}}\simeq 
10^{-6\times 10^{22}}.
\label{eq1}
\end{equation}

The above probability is an incredibly tiny one, but it {\em is not} 
zero. If the total length $l$ of the two-chamber container is 1~meter 
and the mean velocity $\langle v\rangle$ of the particles is 
$\sqrt{\frac{8kT}{\pi m}}$, where $k$ is the Boltzmann's constant, $T$ 
the absolute temperature of the gas and $m$ is the mass of one particle 
($\langle v \rangle$ can be derived from the Maxwell--Boltzmann velocity 
distribution), then the order of magnitude of the average time $\tau$ 
one particle takes to go from chamber $A$ to chamber $B$, or vice-versa, 
is given by

\begin{equation}
\tau=\frac{l}{\langle v\rangle}=\sqrt{\frac{\pi l^2 m}{8kT}}\simeq 
5.6\times 10^{-4}\,
\textrm{seconds},
\label{eq2}
\end{equation}
where for $m$ we chose the mass of the lightest gas molecule (hydrogen 
molecule) and for $T$ the room temperature 298\,K.

One can see $\tau$ as the {\em clock-time} at which the system of 
particles changes its configuration. Hence, the mean time $\langle T 
\rangle$ one has to wait to observe an exceptional occurrence like that 
described above (all particles in chamber $A$) is nearly,

\begin{equation}
\langle T \rangle=\frac{\tau}{P_{A\&B\to A}}\simeq 
10^{6\times 10^{22}}\,\textrm{seconds}.
\label{eq3}
\end{equation}

Note that this time is nearly $10^{6\times 10^{22}}$ times the estimated 
age of the Universe ($\sim 10^{17}\,$seconds) since $\frac{10^{6\times 
10^{22}}\,\textrm{seconds}}{10^{17}\,\textrm{seconds}} \approx 
10^{6\times 10^{22}}$.

Thus, no one will ever have the chance to observe such an occurrence, 
but {\em this does not mean that it is forbidden by the fundamental laws 
of physics}.

For the sake of completeness, let us show why the above situation is a 
violation of the second law of thermodynamics. Let us calculate the 
entropy variation $\Delta S_{A\&B\to A}$ of the two-chamber system soon 
after all particles are in chamber $A$. As is usually done in classical 
thermodynamics, we calculate integral~(\ref{eq0}) along an isothermal 
compression from state {\bf [gas in volume $A\&B$]} to state {\bf [gas 
in volume $A$]} of an ideal gas with equation of state $PV=kNT$ ($k$ is 
the Boltzmann's constant and $N$ the total number of molecules). The 
compression is isothermal and the internal energy $U$ of the gas is 
constant. From the first law of thermodynamics $\delta Q=dU+\delta W$, 
we have $\delta Q=\delta W=pdV=\frac{kN}{V}dV$, and thus

\begin{equation}
\Delta S_{A\&B\to A}=\int_{V_{A\&B}}^{V_A}\frac{\delta Q}{T}=
\int_{V_{A\&B}}^{V_A}\frac{kN}{V}dV=kN\ln\biggl(\frac{V_A}{V_{A\&B}}
\biggr)=-kN\ln2<0,
\label{eq4}
\end{equation}
being $V_{A\&B}=2V_A$.

Actually, consider the above experiment with only 18 molecules in each 
chamber. In such a case eq.~(\ref{eq3}) gives $\langle T \rangle \approx 
1\,$year. This means that every year, on average, this reduced system 
violates the second laws by an amount of $|\Delta S| = k\cdot 36\cdot 
\ln2=3.44\times 10^{-22}\,\frac{\textrm{J}}{\textrm{K}}$, at the most. 
Unfortunately, such a violation could hardly be exploited (to produce 
usable work), not because of its minuteness but because we don't know 
exactly {\em when} this violation happens\footnote{But, somehow, this is 
like saying that the laws of physics {\em forbid} mankind to reach 
Jupiter's satellite Europa simply because we do not have the required 
technology yet.}.

Every other known fundamental laws of physics, like those of Newton's 
mechanics, Einstein's relativity, Maxwell's theory of electromagnetism 
and even the fundamental laws of quantum mechanics (although quantum 
mechanics is intimately linked to an intrinsic probabilistic approach to 
reality, its fundamental laws are not probabilistic) provide an absolute 
and unconditional prescription on how processes should behave in nature. 
For instance, Newton's laws tell us that a body on which no forces work 
undergoes no acceleration; it does not tell us that the body has `a very 
big chance' of undergoing no acceleration. Maxwell's theory tells us 
that two positive charges far removed from any other charge 
distributions will repel each other and how; not that they will repel 
each other `with high probability'. The second law, instead, forbids 
some processes not absolutely, but only with {\em very high 
probability}.

\subsection{Maxwell's Demon: a digression}

\begin{figure}[t]
\begin{center}
\includegraphics[width=5cm]{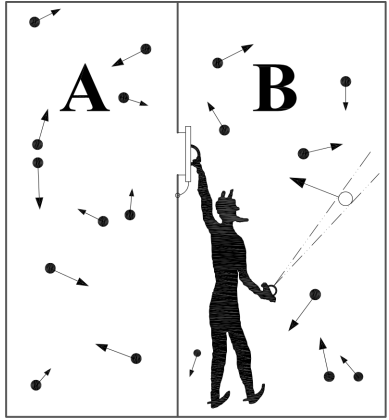}
\caption{A cartoon depiction of Maxwell's demon (adapted from~\cite{le}).}
\label{figdem}
\end{center}
\end{figure}

A two-chamber thought experiment, very similar in the spirit to that 
shown above, dates back to 1867, when J.~C.~Maxwell introduced it for 
the first time to show that the second law of thermodynamics has only a 
{\em statistical validity}. Actually, he made the point more cogent by 
introducing what it is now known as ``Maxwell's Demon'' 
(Fig.~\ref{figdem}). In his own words:

\begin{quote}

[...] if we conceive of a being whose faculties are so sharpened that he 
can follow every molecule in its course, such a being, whose attributes 
are as essentially finite as our own, would be able to do what is 
impossible to us. For we have seen that molecules in a vessel full of 
air at uniform temperature are moving with velocities by no means 
uniform, though the mean velocity of any great number of them, 
arbitrarily selected, is almost exactly uniform. Now let us suppose that 
such a vessel is divided into two portions, A and B, by a division in 
which there is a small hole, and that a being, who can see the 
individual molecules, opens and closes this hole, so as to allow only 
the swifter molecules to pass from A to B, and only the slower molecules 
to pass from B to A. He will thus, without expenditure of work, raise 
the temperature of B and lower that of A, in contradiction to the second 
law of thermodynamics [...].~\cite{ler}

The number of molecules in A and B are the same as at first, but the 
energy in A is increased and that in B diminished, [...] and yet no work 
has been done, only the intelligence of a very observant and 
neat-fingered being has been employed. Or in short if the heat is the 
motion of finite portions of matter and if we can apply tools to such 
portions of matter so as to deal with them separately, then we can take 
advantage of the different motion of different proportions to restore a 
uniform hot system to unequal temperatures or to motions of large 
masses. Only we can't, not being clever enough [...].~\cite{gbe}

[...] I do not see why even intelligence might not be dispensed with and 
the thing be made self-acting. Moral: The 2nd law of Thermodynamics has 
the same degree of truth as the statement that if you throw a tumblerful 
of water into the sea you cannot get the same tumblerful of water out 
again.~\cite{max1}
 
\end{quote}

What is interesting with respect to our previous thought experiment 
(Fig.~\ref{fig2}) is that Maxwell's thought experiment accomplishes a 
violation of the second law that is also an exploitable violation, 
namely one that is able to produce usable work. We have seen with our 
reduced gas-in-two-chambers system above that violation of the second 
law and exploitable violation of the second law are not the very same 
thing.

Maxwell's ``neat-fingered being'' has given rise to an incredibly rich 
literature over the subsequent decades, which is still ongoing and 
stronger than ever. Born as a simple and very effective {\em 
Gedankenexperiment} to elucidate the limits of the second law, almost 
all the subsequent scholars forgot the Maxwell's pristine intention and 
focused their attention entirely on the evil being, trying to exorcise 
it, namely trying to show (prove) the impossibility of the Demon to 
operate in order to save the second law and to preclude the possibility 
of macroscopic exploitation of such a violation (the generation of 
usable work).

Smoluchowski, with his one-way valve~\cite{smol}, and Feynman, with the 
ratchet and pawl analogue~\cite{fey}, introduced a {\em non-sentient} 
version of Maxwell's Demon, using pure physico-mechanical devices 
without the need of an `intelligent being' able to `perceive' 
velocities, `see' paths and `handle' molecules (see the last quotation 
by Maxwell). They have shown that the thermal fluctuations suffered by 
these mechanical devices prevent any anti-entropic action, such the 
sorting of molecules from one vessel to the other.

As a matter of fact, every mechanical device supposed to sort molecules 
must work at the same absolute temperature of the gas; otherwise, its 
action may be ascribed to a possible extraction of work from heat 
reservoirs at different absolute temperatures, like a standard Carnot 
engine, and this does not count as a `regular' second law violation. 
Hence, the mechanical device itself must follow the same canonical 
distribution function associated with the temperature of its immediate 
surrounding.

For instance, in the Smoluchowski one-way valve molecules have an 
average kinetic energy of $\sim kT$ in a given direction, so the 
valve-trapdoor must be sensitive to energies that high, and preferably 
lower energies as well. But the trapdoor has the same temperature as the 
molecules; it is, after all, in contact with them. That means it has 
fluctuations of kinetic energy of the same size as the molecules; that 
is, on the size of $\sim kT$ . The trapdoor must be sensitive to 
energies of order $kT$, and it itself is plagued by fluctuations of 
order $kT$. So it is sensitive to random fluctuations, and there will be 
no correlation between the openings of the trapdoor and the arrival of 
molecules.

Other researches attempted to investigate the {\em sentient} version of 
Maxwell's Demon (probably, the original one), that of intelligently 
operated devices. Szilard and Brillouin argued that in order to achieve 
the entropy reduction, the intelligent being must acquire knowledge on 
molecule's dynamical state (position, velocity) and so must perform a 
measurement. Thus, they argued that the second law would be saved if the 
acquisition of knowledge by the Demon came with a compensating entropy 
cost~\cite{szi,bri}.

In more recent years, some researchers (Bennett, Landauer and followers) 
have claimed that measurements can be performed without entropy costs at 
all. Instead, they focused their attention exclusively on the process of 
information erasure, needed by sentient Demon to operate cyclically. All 
the information gathered and stored by the Demon on the dynamical status 
of the molecules must be first acquired and then necessarily erased in 
order to operate cyclically~\cite{lan,ben,ben2}. According to the 
information erasure school, any kind of sentient Demon is strictly and 
absolutely forbidden to violate the second law by the unavoidable 
entropy cost of the information erasure step, which must be always 
present in order to make the Demon's operation cyclical. This step 
provides the Universe with an entropy increase greater than or equal to 
the alleged entropy reduction operated by the Demon.

Although the connection between physical entropy and information theory 
is now widely recognized, its arguments appear to be either circular 
(themselves typically rely on some version of the second law) or appeal 
to the existence of new profound laws, which have nothing to do with the 
fundamental physical principles (classical and quantum mechanical) that 
govern the behavior of matter~\cite{ear98,ear99,nor05,cal}, and are, in 
the end, a mere recasting of the second law in the lofty formalism of 
information theory: not an explanation of it by the known fundamental 
laws of physics nor a proof of its necessity~\cite{ben03}.

A robust argument against the necessity of information acquisition 
(measurement) and/or memory erasure entropy costs to defeat Maxwell's 
Demon goes as follows. Historically, Szilard~\cite{szi}, 
Bennett~\cite{ben,ben2} and followers have all used the Szilard 
one-molecule heat engine to illustrate their respective point. The 
Szilard heat engine works as follows. Initially the entire volume $V$ of 
a cylinder is available to a single molecule. The first step consists of 
placing a partition into the cylinder, dividing it into two equal 
chambers. In step 2 a Maxwell's Demon determines which side of a 
partition the molecule is on, and records this result. In step 3 the 
partition is replaced by a piston, and the recorded result is used to 
couple the piston to a load upon which work $W$ is done. The gas 
pressure move the piston to one end of the container, returning the gas 
volume to its initial value, $V$. In the process the one-molecule gas 
has energy $Q=W$ delivered to it via heat from a constant temperature 
heat reservoir. After step 3 the gas has the same volume and temperature 
it had initially. The heat bath, which has transferred energy to the 
gas, has lower entropy than it had initially. Without some other 
mechanism, the second law has been violated during the cyclic process.

Szilard and followers (Brillouin\footnote{In particular, 
Brillouin~\cite{bri} mathematically addressed in explicit way the 
original form of Maxwell's Demon, namely that of a ``neat-fingered 
being'' able to actually see individual molecules. He showed that in 
order to see the single molecule the Demon should use a (black-body) 
radiation more energetic (higher temperature) than the black-body 
radiation of the gas and environment, thus generating a compensating 
entropy increase.} and Gabor being the most representative ones) 
suggested that one may reasonably assume that an amount of entropy is 
generated during the measurement process by the Demon (in order to know 
which side of the partition the molecule is on) that restores 
concordance with the second law. Bennett and followers, instead, argued 
that the Demon must erase its record on the position of the molecule in 
order to make the whole process cyclic. Thus, they associated with the 
erasure step an entropy increase no less than the entropy reduction 
operated by the Demon.

\begin{figure}[t]
\centerline{\includegraphics[width=6cm]{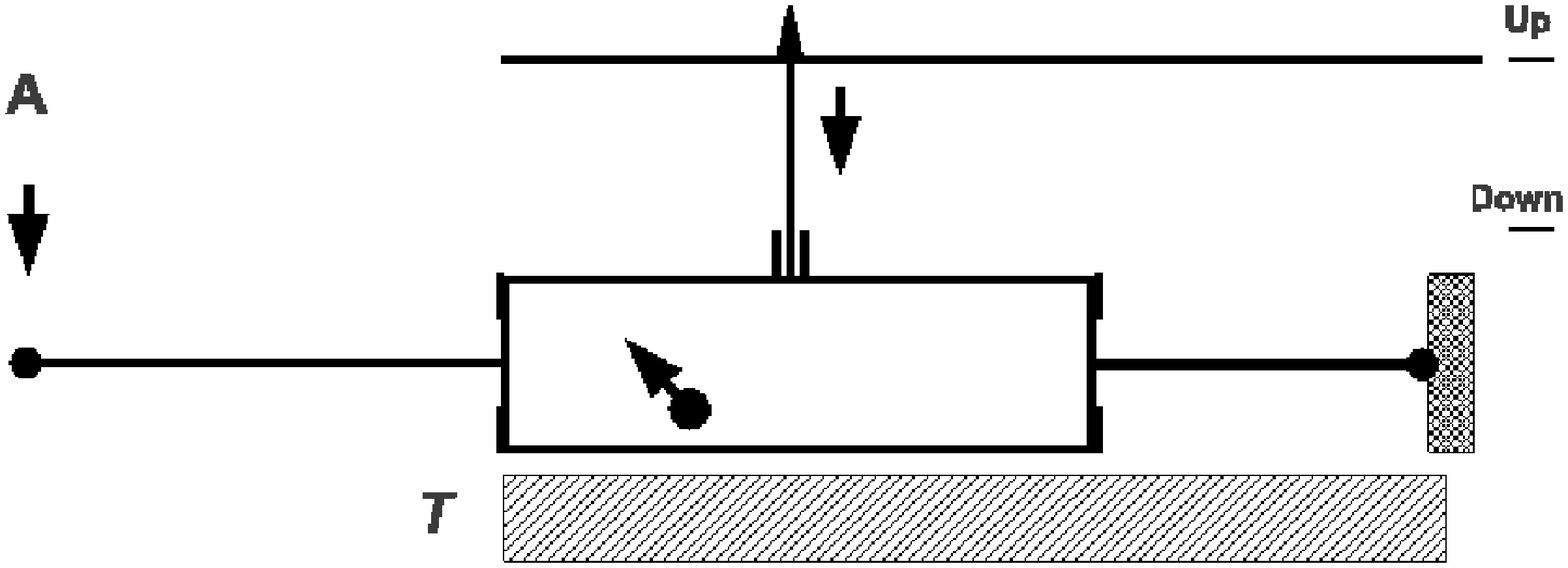}}
\vspace{0.5cm}
\centerline{\includegraphics[width=6cm]{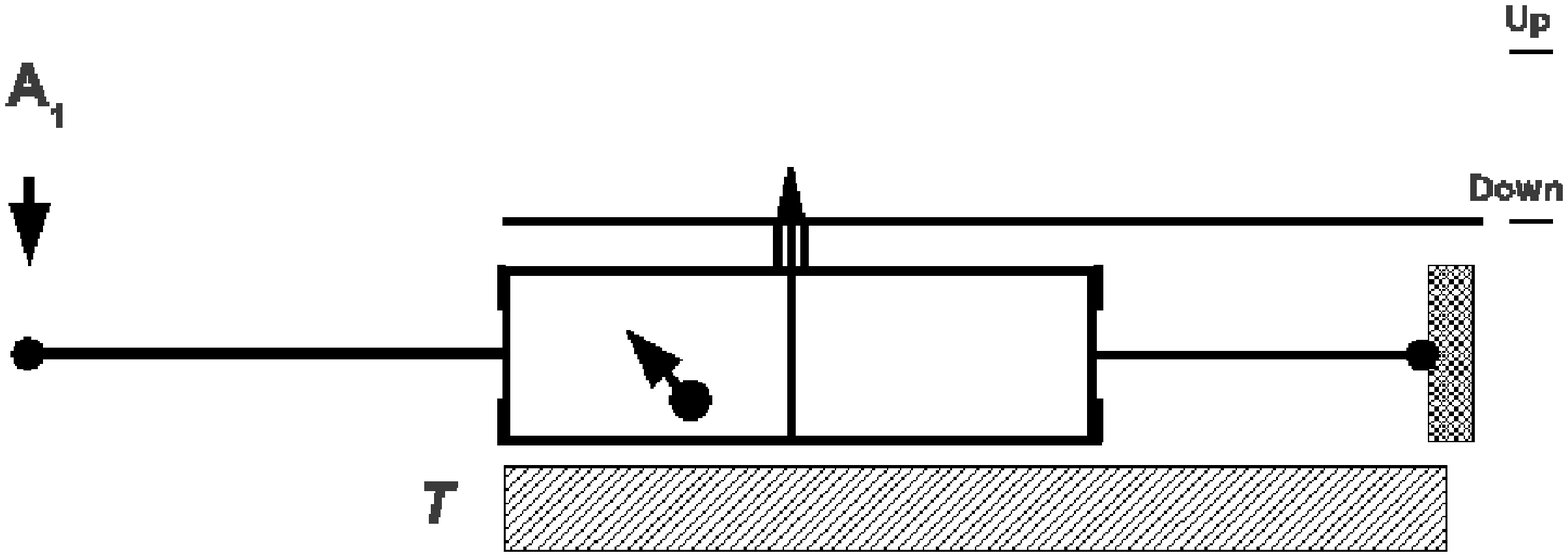}}
\vspace{0.5cm}
\centerline{\includegraphics[width=6cm]{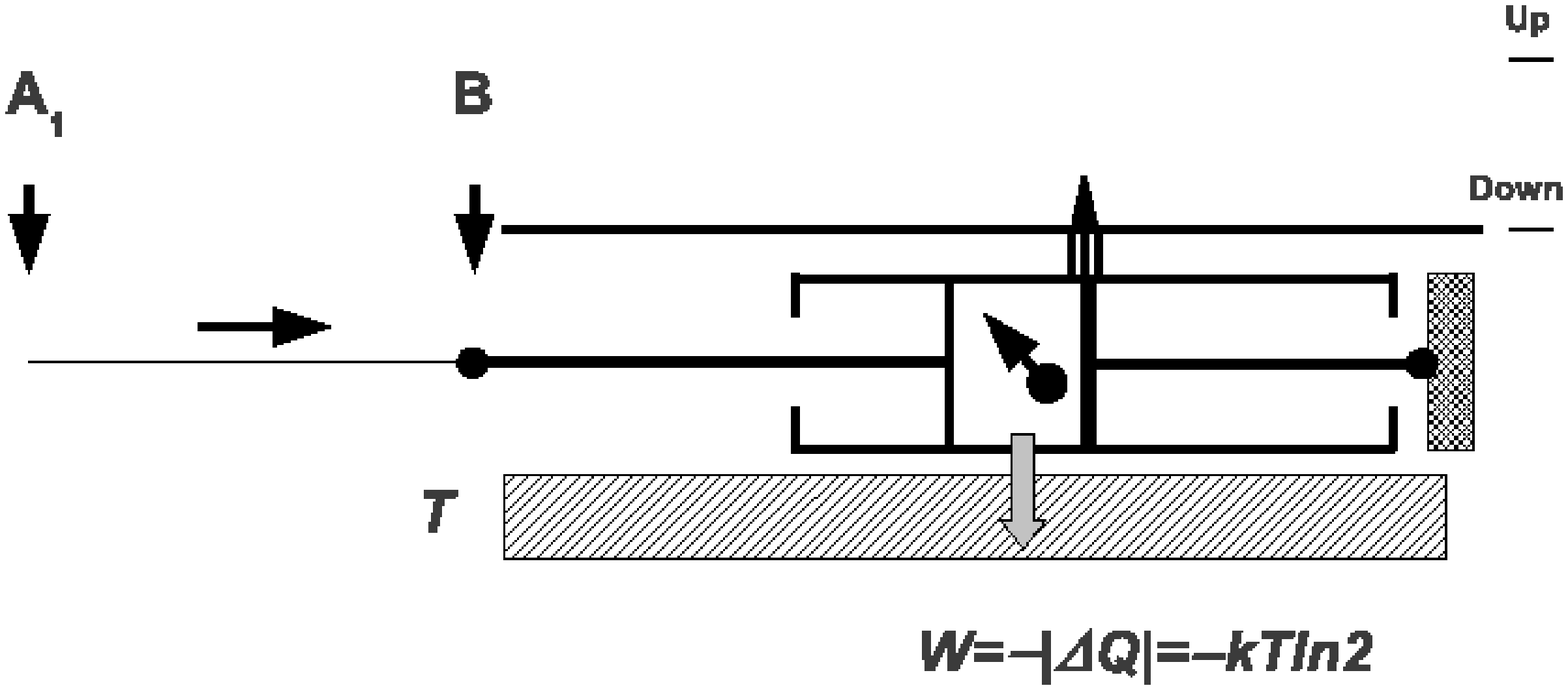}}
\vspace{0.5cm}
\centerline{\includegraphics[width=6cm]{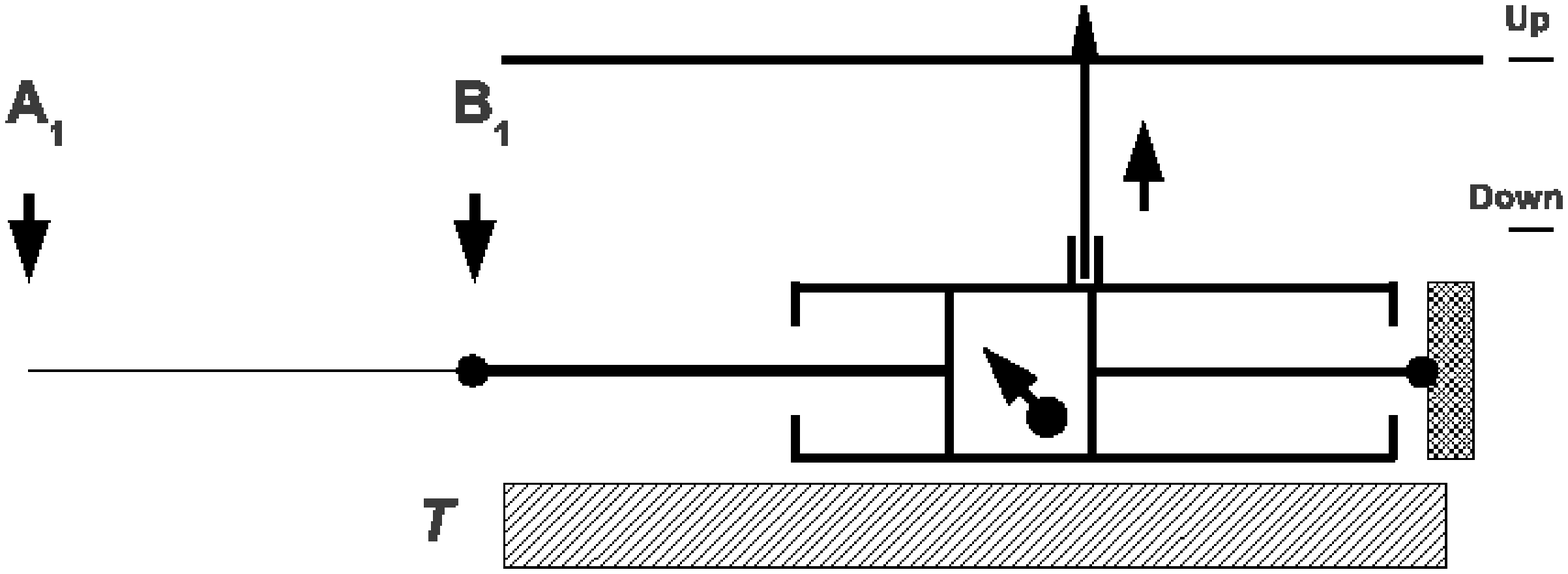}}
\vspace{0.5cm}
\centerline{\includegraphics[width=6cm]{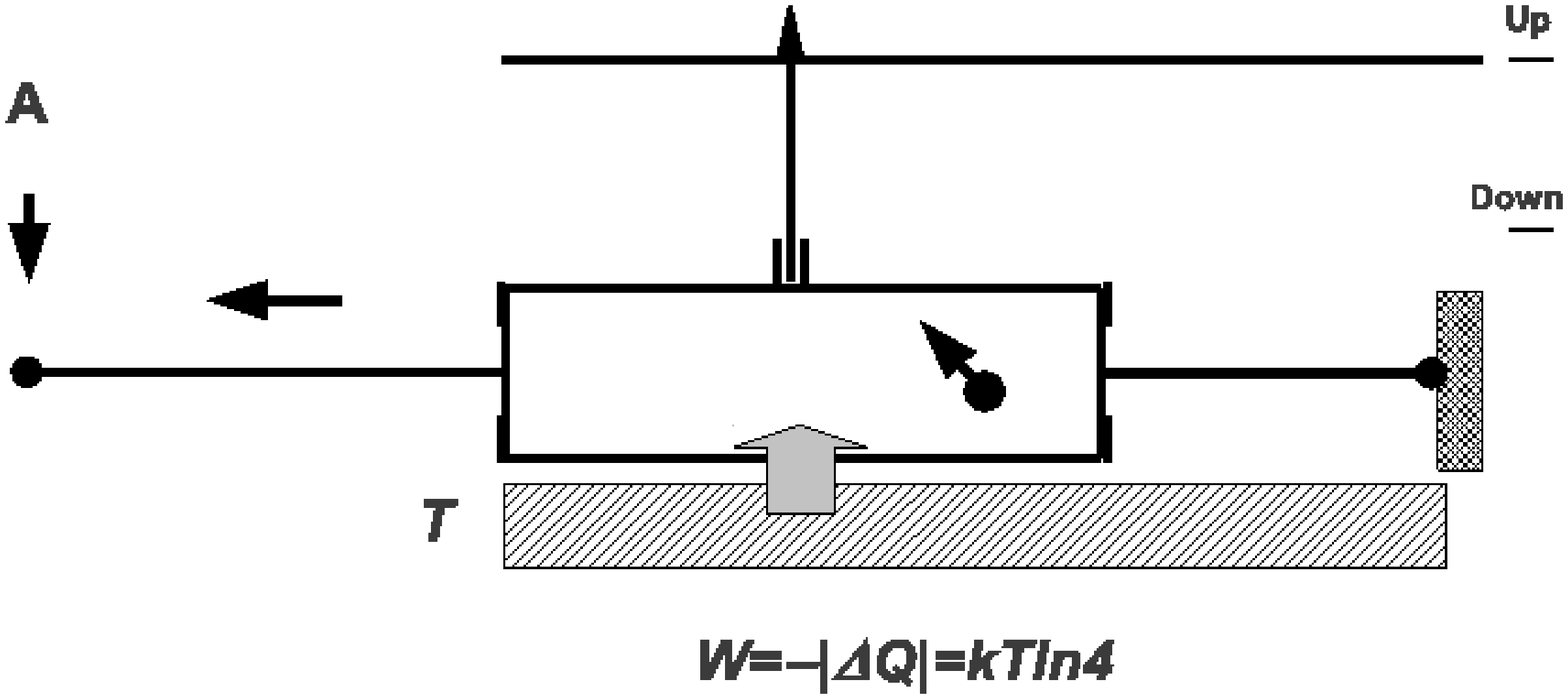}}
\caption{The modified Szilard's heat engine described in the text.}
\label{fig4b}
\end{figure}

As a matter of fact, it is not difficult to devise a modified Szilard's 
heat engine which cyclically works without the need of information 
acquisition and/or memory erasure. Such an engine is shown in 
Fig.~\ref{fig4b}. It is made of a movable cylinder and two pistons (the 
left one movable and the right one fixed). There is also a partition 
that can be lowered in the middle of the cylinder and that can slide 
horizontally on a lowering rod without friction as the cylinder moves. 
The insertion of the partition involves no work or heat. All the 
mechanical parts are thought without friction, as has been done 
extensively in the literature on the subject (more on this later). 
Initially the entire volume $V$ of a cylinder is available to a single 
molecule (step A). The behavior of the molecule is described by the 
equation of state $PV=kT$. Then, the partition is lowered into the 
cylinder, dividing it into two equal chambers (step A$_1$). The molecule 
is trapped in one of these two chambers.

Then, the movable piston is pushed infinitesimally slowly (reversibly) 
to position B and the one-molecule gas undergoes an isothermal 
compression from $V/2$ to $V/4$. The work $W_{A_1\to B}$ externally done 
to the gas is equal to $kT\ln 2$, which is also equal to the heat 
transferred by the gas to the heat reservoir at temperature $T$. Note 
that this part of the cycle is independent of which side of the 
partition the molecule is on at step A$_1$, hence we do not need any 
information acquisition (with subsequent memory erasure). The final 
position of the movable piston is always at point B, no matter if the 
molecule is in the right or in the left chamber after step A$_1$. 
Besides, the movement of the partition can be mechanically coupled to 
the cyclic movement of the movable piston, and thus without any need of 
information acquisition and/or memory erasure to operate the partition 
itself.

One may complain that the compression procedure depends on whether the 
molecule is trapped on the left or right. Namely, if the molecule is on 
the left, the piston moves the whole cylinder first, with its action on 
the cylinder mediated by the gas pressure. If the molecule is on the 
right, the piston moves in unimpeded to contact the partition and then 
compresses the gas. Since the two processes appear to be slightly 
different, one may wonder that in order to operate the device you have 
to know which conditions is at hand. This would mean measurement and/or 
memory erasure. Under a more careful analysis one can easily see that 
the two processes are not different at all. In both cases there is a 
first phase where the device moves unimpeded until the right piston, if 
the molecule is on the left, or the left piston, if the molecule is on 
the right, touches the partition (from step A$_1$ to the midpoint 
between A$_1$ and B), and then a second phase in which there is the true 
gas compression (from the midpoint between A$_1$ and B to step B). These 
two phases are physically perceived always in the same way by who/what 
operates the device: the first half of the compression is always equally 
`loose', no matter where the molecule is at the beginning of the 
process, while the second half is the true gas compression.
 
At step B$_1$, the partition is raised and the cycle is completed with 
an isothermal expansion from $V/4$ to $V$ (with movable piston again in 
position A). Now, the work $W_{B_1\to A}$ made by the gas to the 
environment is equal to $kT\ln 4$, which is also equal to the heat 
transferred by the heat reservoir at temperature $T$ to the gas. The net 
work output $W_n$ over any cycle is then equal to $W_{B_1\to 
A}-W_{A_1\to B}=kT(\ln 4 -\ln 2)=kT\ln 2$. Moreover, the entropy 
variation $\Delta S$ of the entire system (engine + reservoir) is equal 
to $-k\ln 2$.

If we want to save the second law in the above scheme, then some other 
mechanisms must come into play to prevent the modified Szilard engine 
from operating. For instance, thermal fluctuations surely afflict the 
mechanical parts of the engine (pistons, partition and so 
on)~\cite{ear99}. The pistons must be sensitive to energy of the order 
of $kT$, the mean energy of the molecule, and they themselves are 
plagued by fluctuations of order $kT$, like the Smoluchowski one-way 
valve. Actually, if there were no friction, then the device could 
operate even with arbitrarily massive pistons, partition and cylinder 
(massive means not instantaneously sensitive to energy of the order of 
$kT$). As a matter of fact, without friction even the tiny kick of a 
single molecule can move a massive piston/cylinder (conservation of 
linear momentum). But friction cannot be eliminated, even ideally, since 
thermal fluctuations of the matter along the contact points between the 
pistons' edge and the cylinder's walls originate an unavoidable friction 
force that is surely greater than the force imparted by the molecule to 
the pistons.

But, if such effects afflict our modified Szilard's engine, then the 
same effects must afflict the original Szilard's engine, being both 
engines mechanically similar. Hence, the appeal to information 
acquisition and/or memory erasure entropy costs to defeat the Maxwell's 
Demon in the instantiation of the original Szilard's engine is 
superfluous. On the other hand, if information acquisition and/or memory 
erasure entropy costs are strictly necessary to defeat original 
Szilard's engine, then this means that no other mechanisms are able to 
prevent its operation. But this last thing would necessarily apply also 
to our modified Szilard's engine. Thus, our engine would surely violate 
the second law, since measurement and memory erasure, with their 
associated entropy costs, do not apply to it, as we saw above. As a 
logical consequence, measurement and memory erasure entropy costs are 
again unnecessary to defeat Maxwell's Demon, this time in the 
instantiation of our modified Szilard's engine.

As a conclusion, the appeal to measurement and memory erasure entropy 
costs made by Szilard and Bennett within the original Szilard's engine 
appears to be an arbitrary choice rather than a necessity in defeating 
the Szilard's Demon.

Probably the true reason why non-sentient Demon cannot operate, namely 
cannot macroscopically violate the second law and create usable work, is 
the ubiquity of thermal fluctuations and friction in the physical 
matter, the matter which inevitably constitutes both gas and every 
passive device conceived to sort molecules in the gas-in-two-chambers 
scheme.

Fluctuations make the non-sentient Demon ineffective and ultimately 
Maxwell's thought experiment of the two-chamber vessel with a 
non-sentient Demon becomes equivalent to our thought experiment 
(Fig.~\ref{fig2}) of the two chambers connected by a wide open hole: 
macroscopic violation of the second law can be possible {\em only} by 
macroscopic statistical fluctuations of molecules between chambers, with 
or without a sorting Demon.

In addition, every sentient Demon (for instance, like that of Brillouin) 
which in order to operate needs to acquire information on the molecule 
(or even to erase memory) necessarily must release (exchange) energy to 
the gas and the environment: this is equivalent to a Demon which 
performs work to the system. It is not a canonical Maxwell's Demon, 
which operates ``without expenditure of work'', in Maxwell's own words. 
Thus, there is the strong feeling that every sentient Demon is doubly 
ineffective in violating the second law in the gas-in-two-chambers 
scheme: firstly, because every mechanical part of it, which has to be 
`picometric' in order to deal with single molecules, is unavoidably 
plagued by thermal fluctuations\footnote{In Feynman's own words: `If we 
assume that the specific heat of the demon is not infinite, it must heat 
up. It has but a finite number of internal gears and wheels, so it 
cannot get rid of the extra heat that it gets from observing the 
molecules. Soon it is shaking from Brownian motion so much that it 
cannot tell whether it is coming or going, much less whether the 
molecules are coming or going, so it does not work.'. Besides, in a 
recent work John Norton provides a general result which seems to 
rigorously show the unavoidability of such a limitation~\cite{wai}.} and 
friction; second, because every energy exchange with the gas required 
by the measurement process (or even by the memory erasure) may imply a 
further entropy increase.

\subsection{Back to the second law}

The critical evaluation of the literature on Maxwell's Demon and, to 
some extent, of Maxwell's Demon itself are not the main goal of this 
paper; the interested reader is referred to \cite{ler}, \cite{ear98}, 
\cite{ear99}, \cite{mar}, \cite{gij} and the references therein.

Rather, our interest is mainly in the epistemological significance of 
the gas-in-two-chambers scheme (with or without Maxwell's Demon) for the 
status of the second law.

Given the above, the only logically tenable, legitimate and more basic 
inference that can be drawn from the gas-in-two-chambers thought 
experiment (that of Maxwell but, above all, that depicted in 
Fig.~\ref{fig2} and described before), is {\bf not} that the probability 
of a macroscopic and exploitable violation of the second law is {\em 
always} extremely small (practically zero) and thus the second law is 
safe, but that:

\begin{itemize}

\item[i)] the second law is {\bf not a necessary law}. There aren't 
known fundamental laws of physics which absolutely forbid its violation 
and thus it can be macroscopically violated in principle. None of 
Maxwell's Demon exorcisms provides basic principles and fundamental laws 
of physics able to absolutely forbid the violation of the second law.
 There is no exorcism that is not attributable in the end to thermal 
fluctuations and friction convincingly and beyond a reasonable doubt;

\item[ii)] the probability of a macroscopic and exploitable violation of 
the second law is extremely small {\bf if} one uses the 
gas-in-two-chambers scheme or analogues, with or without a sorting 
Demon. As a matter of fact, thermal fluctuations make every 
gas-in-two-chambers scheme with a sorting (sentient or non-sentient) 
Demon equivalent to a gas-in-two-chambers scheme with a wide open hole 
between the two chambers. Thus, macroscopic violations of the second law 
are possible only by macroscopic statistical fluctuations of molecules 
between chambers. And we know that this is statistically highly 
improbable.

\end{itemize}

From Maxwell's and our thought experiment one cannot definitively infer 
that the second law cannot be macroscopically violated by schemes {\em 
different} form the gas-in-two-chambers ones, those for instance not 
involving gas, liquid or solid atoms and molecules in thermal 
equilibrium (whose behaviour is described by the canonical 
distribution).

For what concerns the gas-in-two-chambers scheme described before 
(Fig.~\ref{fig2}), the following summary inference chart holds:

{\small
\begin{displaymath}
\begin{array}{lcl}
\textrm{1) Gas spontaneous macroscopic} & \Rightarrow & 
\textrm{Macroscopic violation of the second}\\
\textrm{compression in the gas-in-two-} & & \textrm{law}\\
\textrm{chambers scheme} & &\\
& & \\
& \textrm{But} & \\
& & \\
\textrm{2) Practical impossibility of gas} & \nRightarrow & 
\textrm{Absolute macroscopic non-violability}\\
\textrm{spontaneous macroscopic compression} & & \textrm{of the second law}\\
\textrm{in the gas-in-two-chambers scheme} & &\\ 
& & \\
\textrm{3) Macroscopic violation of the second} & \nRightarrow & \textrm{Real possibility of gas spontaneous}\\
\textrm{law} & & \textrm{macroscopic compression in the gas-}\\
 & & \textrm{in-two-chambers scheme, \em which is}\\
 & & \textrm{\em actually anything but probable}
\end{array}
\end{displaymath}
}

Namely, the inability of the gas-in-two-chambers scheme (with or without 
Maxwell's Demon) to macroscopically violate the second law is 
\emph{logically} uncorrelated with the actual possibility of second law 
macroscopic violation.

The inference 3) has been explicitly added since sometime people are 
overwhelmed by the logical fallacy that if the second law can be somehow 
macroscopically violated, then this would automatically imply that gas 
spontaneous compressions in the gas-in-two-chambers scheme would be 
actually possible. Then, with a sort of `inverted logic', they argue 
that being such compressions statistically highly improbable then the 
second law cannot be macroscopically violated. These two facts, as 
showed in point 3) of the inference chart, are uncorrelated in such an 
inference direction.

What we are suggesting is that Boltzman's principle of statistical 
entropy increase\footnote{A system approaches equilibrium because it 
evolves from states of lower toward states of higher probability, and 
the equilibrium state is the state of highest probability.} (well 
represented by the high improbability of spontaneous gas compression) 
and the macroscopic violation of the second law of thermodynamics can be 
both true or, better, are not mutually exclusive (see also~\cite{dun}). 
By the way, Versteegh and Dieks, in a very interesting paper on the 
Gibbs paradox and the distinguishability of identical 
particles~\cite{vd}, point out that the entropy concept in 
thermodynamics is not completely identical to that in statistical 
mechanics.

While {\em necessity} by fundamental physics principles would mean 
strict inviolability, as far as the known fundamental laws of physics 
are valid, a possible non-necessary (contingent) nature of the second 
law, as is clear from the above considerations, leaves the door open for 
its violability. Obviously, contingency is a necessary but not 
sufficient condition for violability. Given the actual status of the 
second law, research aiming at the study of its violability appears to 
be worthwhile.

\section{The quest for violation}

Over the past 30 years, an unparalleled number of challenges has been 
proposed against the status of the second law. During this period, more 
than 50 papers have appeared in the refereed scientific 
literature~\cite{cs}. Moreover, during the same period three 
international conferences on the limit of the second law were also 
held~\cite{cong1, cong2, cong3}.

Obviously, not all the scholars are willing to acknowledge a respectable 
status to this line of research. For instance, Gyftopoulos \& Beretta 
wrote:

\begin{quote}
If no challenges have been proven valid [so far], what is the motivation 
for pursuing exploratory research to prove that the second law is 
invalid? To put our question differently, why people interested in 
exploratory research do not try to prove that the solar system is 
neither geocentric nor heliocentric? Similarly why researchers do not 
try to prove that, in the realm of its validity, Newton's equation of 
motion is not correct?~\cite{gb}
\end{quote}

The straight answer to this question is that, as argued before, both 
Newton's laws of mechanics and the heliocentric theory hold a different 
(epistemological) status with respect to the second law of 
thermodynamics.

The general class of recent challenges spans classical/standard 
\cite{go1, go2, go3, go4, de1, de2, de3, de4, cro1, cro2, cro3, cro4, 
dab1, dab2}, plasma \cite{she1, she2, she3}, chemical \cite{she4, she5, 
she6, she7}, gravitational \cite{she11, she12, she13, she14}, solid 
state~\cite{she8, she9, she10} and quantum physics \cite{cap1, cap2, 
cap3, cap4, cap5, cap6, cap7, cap8, cap9, cap10, cap11, cap12, cap13, 
cap14, cap15, cap16, an1, an2, an3, an4, an5, kee1, kee2, kee3, nik1, 
nik2, be1, be2}. Some of these approaches appear immune to standard 
second law defenses and several of them account laboratory corroboration 
of their underlying physical processes. Others, mainly some 
classical/standard and gravitational challenges, have been criticized 
and/or proved faulty beyond any reasonable doubt (see, for 
example~\cite{mot, cha, whe}).

A thorough description of all these challenges is a quite hard task to 
accomplish and it is beyond the scope of this paper. The interested 
reader may find a very detailed review in \v C\'apek \& 
Sheehan~(2005)~\cite{cs} and Sheehan~(2008)~\cite{shecomp}.

From the point of view of direct laboratory testability, the more 
promising challenges appear to be the solid state ones~\cite{she8, she9, 
she10}, together with the challenge posed by thermo-charged 
capacitors~\cite{dab1, dab2}.

\subsection{Solid state challenges}

\begin{figure}[t]
\begin{center}
\includegraphics[height=6cm]{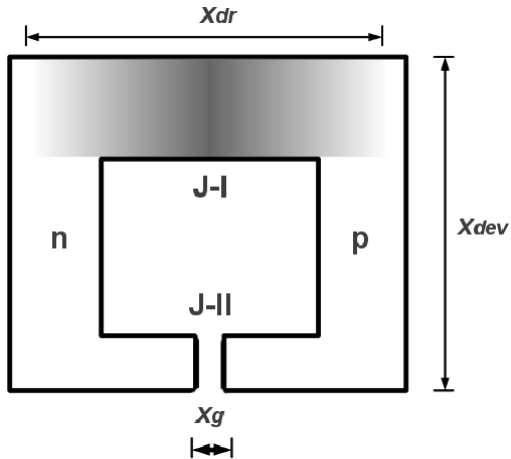}
\end{center}
\caption{The core of the solid state challenge. Adapted from~\cite{she8}.}
\label{fig3}
\end{figure}

Since 2002, two types of solid state devices have been proposed 
\cite{she8, she9, she10, cs} that basically utilize the electric field 
energy of an open-gap {\em n-p} junction. They are based on the cyclic 
electro-mechanical discharging and thermal recharging of the 
electrostatic potential energy intrinsic to the depletion region of a 
standard solid state {\em n-p} junction. The core of their functioning 
is the shaped junction depicted in Fig.~\ref{fig3}.

It consists of two symmetric horseshoe-shaped pieces of {\em n}- and 
{\em p}-se\-mi\-con\-duc\-tor facing one another. At Junction I (J-I), 
the {\em n}- and {\em p}-regions are physically connected, while at 
Junction II (J-II) there is a vacuum gap whose width $x_g$ is small 
compared to the scale lengths of either the depletion region $x_{dr}$ or 
the overall device $x_{dev}$; namely, $x_g\ll x_{dr}\sim x_{dev}$. All 
the scale lengths are in the {\em micro-}, {\em nano-metric} realm.

As is well known from solid state physics, a built-in potential $V_{bi}$ 
forms across the junction J-I, whose numerical value depends on the 
doping characteristics of the two regions (concentrations of donors and 
acceptors, intrinsic carrier concentration) and on the environmental 
temperature (in the present case, room temperature). Its value can be 
estimated analytically.

This potential is the result of charge diffusion across J-I as soon as 
the the two materials are physically joined. The depletion region is 
thus the region where, at equilibrium, a balance between bulk 
electrostatic and diffusive (thermally driven) forces is attained.
 
It is then claimed that an electric field must exist also in J-II. 
According to~\cite{she8, she9, she10, cs}, the existence of an electric 
field in the J-II gap at equilibrium can be established either via 
Kirchhoff's loop rule (conservation of energy) or via path-independence 
$\oint \textrm{\bf E} \cdot d\textrm{\bf l} = 0$. It is argued as 
follows. Consider a vectorial loop threading the J-I depletion region, 
the bulk of the device in Fig.~\ref{fig3}, and the J-II gap. Since the 
built-in electric field in the J-I depletion region is unidirectional, 
there must be a second electric field somewhere else along the loop to 
satisfy $\oint \textrm{\bf E} \cdot d\textrm{\bf l} = 0$. An electric 
field elsewhere in the semiconductor bulk (other than in the depletion 
region), however, would generate a current, which contradicts the 
assumption of equilibrium. Therefore, by exclusion, the other electric 
field must exist in the J-II gap. Kirchhoff's loop rule establishes the 
same result. Conservation of energy demands that a test charge conveyed 
around this closed path must undergo zero net potential drop; therefore, 
to balance $V_{bi}$ in the depletion region, there must be a 
counter-potential somewhere else in the loop. Since, at equilibrium, 
away from the depletion region in the bulk semiconductor there cannot be 
a potential drop (electric field)- otherwise there would be a 
non-equilibrium current flow, contradicting the assumption of 
equilibrium - the potential drop must occur outside the semiconductor; 
thus, it must be expressed across the vacuum gap J-II.

Because the J-II gap is narrow ($x_g\ll x_{dr}$) and the built-in 
potential is discontinuous (due to the vacuum gap), there can be large 
electric fields there, which can be much greater than in the J-I 
depletion region. As a matter of fact, one can estimate the relative 
magnitude as follows: the J-II electric field is $|\textrm{\bf 
E}_{\textrm{\bf J-II}}|\simeq \frac{V_{bi}}{x_g}$, while the average 
magnitude of the field in J-I is $|\textrm{\bf E}_{\textrm{\bf 
J-I}}|\simeq \frac{V_{bi}}{x_{dr}}$, thus their ratio scales as 
$\frac{|\textrm{\bf E}_{\textrm{\bf J-II}}|}{|\textrm{\bf 
E}_{\textrm{\bf J-I}}|} \sim \frac{x_{dr}}{x_{g}}\gg 1$.
 
Through a mathematical treatment of the device, it has been 
shown~\cite{she8, cs} that if some provisos on $x_{g}$ and $x_{dr}$ are 
met, then the electrostatic potential energy in J-II (electrostatic 
energy density times gap volume) is much greater than that in the entire 
depletion region J-I. Moreover, if the open gap J-II is switched closed 
(thus becoming a second J-I junction), then such an excess energy is 
positively released. Most of the free electronic charges on each gap 
face disperse through and recombine in the J-II bulk.

It is clear that if such a release can be made cyclical through an 
electro-mechanical nano-apparatus, then this kind of device can exploit 
the thermally driven diffusion across J-I to produce usable work. 
Namely, it appears to violate the second law of thermodynamics in the 
Kelvin--Planck formulation.

Two kinds of such an electro-mechanical apparatus have been proposed 
and modeled so far (both analytically and numerically), one which uses a 
Linear Electrostatic Motor (LEM)~\cite{she8, cs}, and the other 
using an Hammer and Anvil analogue~\cite{she9, she10, shecomp,cs}.

Although the existence of an intense electric field in J-II has been 
recently put into question on the basis of some heuristic and 
theoretical arguments (which do not appeal circularly to the validity of 
the second law \cite{dab3}), micro-metric partial hammer and anvil 
prototypes have been fabricated and are currently undergoing laboratory 
tests. The authors report that preliminary results appear to be 
positive~\cite{shesd}.

\subsection{Thermo-charged capacitors}

Thermo-charged capacitors are vacuum capacitors spontaneously charged 
harnessing the heat absorbed from a single thermal reservoir at room 
temperature~\cite{dab1, dab2}.

\begin{figure}[t]
\begin{center}
\includegraphics[height=6cm]{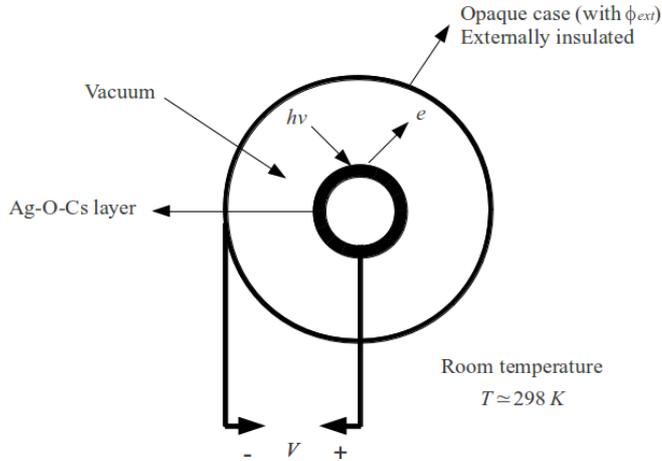}
\end{center}
\caption{Scheme of the thermo-charged spherical capacitor. Adapted 
from~\cite{dab1}.}
\label{fig4}
\end{figure}

In Fig.~\ref{fig4} a sketched section of a vacuum spherical 
thermo-charged capacitor is shown. The outer sphere has radius $b$ and 
it is made of metallic material with relatively high work function 
($\phi_{ext} > 1\,$eV). The inner sphere has radius $a$ and it is made 
of the same conductive material as the outer one, but it is coated with 
a layer of semiconductor Ag--O--Cs, which has a relatively low work 
function ($\phi_{in}\lesssim 0.7\,$eV). In such a case the two 
thermionic fluxes, from each plate toward the other one, are different, 
the latter being greater than the former, at least at the beginning of 
the charging process. The capacitor is shielded by a case and put at 
room temperature. The case is opaque to every environmental 
electromagnetic disturbance (natural and man-made e.m.~waves, cosmic 
rays and so on) in order to avoid spurious {\em photo}-electric 
emission. Moreover, the outer plate is externally insulated, in order to 
prevent its outward thermionic emission and the inter-plate space is 
under extreme vacuum (UHV).

All the electrons emitted by the inner sphere, due to thermionic 
emission at room temperature, are collected by the outer (very low 
emitting) sphere, creating a macroscopic difference of potential $V$. At 
first, such a process is unbalanced, the flux from the inner sphere 
being greater than that from the outer sphere, but later, with the 
increase of potential $V$, the inward and the outward {\em effective} 
flux tend to balance each other exactly.

\begin{figure}[t]
\centerline{\includegraphics[width=8cm]{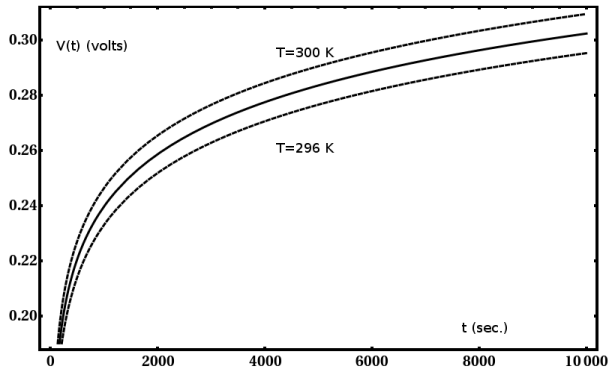}}
\caption{Thermo-charging profiles $V(t)$ for a spherical capacitor with 
$\phi_{in}=0.7\,$eV, $\phi_{ext}=4.0\,$eV, $b=0.2\,$m, $T=298\,$K, and 
conservative values of the mean quantum efficiencies 
$\overline{\eta}_{in}=10^{-5}$ and $\overline{\eta}_{ext}=1$. Charging 
profiles for $T=300\,$K and for $T=296\,$K are also shown. Adapted 
from~\cite{dab2}.}
\label{fig5}
\end{figure}

It has been shown~\cite{dab2} that, under conservative conditions, the 
differential equation which governs the process of thermo-charging is,

\begin{equation}
\frac{dV(t)}{dt}= \frac{\pi e b}{2\epsilon_0
c^2}\biggl(\frac{kT}{h}\biggr)^3 \Biggl(\overline{\eta}_{in}\int_{\frac{eV(t) +
\phi_{in}}{kT}}^\infty \frac{x^2 dx}{e^{x}-1} - 
4\overline{\eta}_{ext}\int_{\frac{\phi_{ext}}{kT}}^\infty
\frac{x^2 dx}{e^x-1}\Biggr),
\label{eq6}
\end{equation}
where $\epsilon_0$ is the vacuum permittivity, $c$ is the speed of light, 
$e$ is the electronic charge, $k$ is Boltzmann's constant, $h$ is 
Planck's constant, $\overline{\eta}_{in}$ is the mean quantum efficiency 
of thermionic material Ag--O--Cs and $\overline{\eta}_{ext}$ is the mean
quantum efficiency of the metallic outer sphere. It can be proved that 
the radius of the inner sphere $a$ must be equal to half the radius of 
the outer sphere $b$ in order to have the maximum charging efficiency 
(faster charging rate). In Fig.~\ref{fig5} the charging profiles for a 
20\,centimeters capacitor are shown.

The charging process is a quite straightforward physical mechanism and 
it appears almost unproblematic. However, during the charging of the 
device the inner thermionic sphere substantially absorbs heat from the 
environment and releases this energy to the thermionic electrons. Such 
electrons fly to the outer sphere and impinge on it with non-zero 
velocity (since a non-zero fraction of them gathers their kinetic energy 
from the high energetic tail of the Planck distribution of black-body 
radiation). When they impinge on the outer sphere, they release their 
kinetic energy substantially heating the outer sphere. Thus, we are 
facing a spontaneous process involving an isolated system at uniform 
temperature (capacitor + environment), in which a part of the system 
(the inner sphere of the capacitor) absorbs heat at temperature $T$ and 
transfers a fraction of this heat to the other part of the system (the 
outer sphere) at the same temperature. This seems to macroscopically 
violate the second law of thermodynamics in the Clausius formulation. In 
Maxwell's own words:

\begin{quote}
One of the best established facts in thermodynamics is that it is 
impossible in a system enclosed in an envelope which permits neither 
change of volume nor passage of heat, and in which both the temperature 
and the pressure are everywhere the same, to produce any inequality of 
temperature or of pressure without the expenditure of work.~\cite{max3}
\end{quote}

As a matter of fact, if $Q_i$ is the energy absorbed by the inner sphere 
from the environment, $U$ is the energy stored in the electric field 
between the spheres ($U=\frac{1}{2}CV^2$, where $C=\frac{4\pi\epsilon_0 
ab} {b-a}$ is the capacitance of the spherical capacitor), and $Q_f$ is 
the energy transferred through the flying electrons to the outer sphere 
as heat (according to the first law of thermodynamics $Q_f+U=Q_i$, thus 
$Q_i>Q_f$), then the Clausius entropy variation of the whole system, as 
rough estimate, amounts to

\begin{equation}
\Delta S_{tot}\simeq-\frac{Q_i}{T}+ \frac{Q_f}{T}<0. 
\label{eq7}
\end{equation}

In order to make the above result more striking, let us consider the 
following analogue in classical thermodynamics/mechanics: a boulder of 
mass $m$ rests at the bottom of a valley, below a hill of height $h$, 
all the system at constant temperature $T$. Suddenly, the boulder 
spontaneously absorbs an amount $Q_1$ of heat (energy) from the 
environment and spontaneously starts to climb the hill at decreasing 
velocity (since the initial kinetic energy is gradually transformed into 
gravitational potential energy). Near the top of the hill the boulder 
hits a sort of wall and then stops. The friction experienced during the 
hit against the wall lets the boulder release to the environment an 
amount $Q_2$ of heat, obviously smaller than $Q_1$. According to the 
first law of thermodynamics we have: $Q_1-Q_2=mgh$, where $mgh$ is the 
gravitational potential energy variation of the boulder from the valley 
to the top of the hill.

Now, the total Clausius entropy variation is:

\begin{equation}
\Delta S_{tot}=-\frac{Q_1}{T}+\frac{Q_2}{T}=-\frac{mgh}{T}<0.
\label{eq8}
\end{equation}

The behavior of the boulder-environment system is practically the same 
as that of our electrons-environment system, and it is undoubtedly 
puzzling from the point of view of the second law of thermodynamics.

Furthermore, the behavior of the electrons just after the emission from 
the Ag--O--Cs coating is governed by the mechanical/ballistic laws of 
motion and not by the canonical distribution which describes systems in 
thermal equilibrium, $p(\mathbf{x},\mathbf{p})=\frac{e^{-E(\mathbf{x}, 
\mathbf{p})/kT}}{Z}$, where $E(\mathbf{x},\mathbf{p})$ is the energy of 
the system, $Z$ is the normalizing partition function, and the 
multi-component $\mathbf x$ and $\mathbf p$ are generalized 
configuration and momentum coordinates. Hence, the randomizing 
(disruptive) effect of thermal fluctuations for cases described 
in~\cite{wai} does not appear to apply here.

Concerning the exploitability of such an alleged violation, it has been 
shown in~\cite{dab2} that the potential drop $V$ reached during the 
thermo-charging process is rapidly transferred to the terminal leads of 
the capacitor (Fig.~\ref{fig4}).

The junction between Ag--O--Cs coating material and the inner metallic 
sphere is a Schottky junction and behaves like a rectifying diode. In 
principle, such a rectifying behavior could prevent the potential drop 
$V$ attained within the two concentric spheres from reaching the 
terminal leads, and thus could forbid any exploitation.

As a matter of fact, any real rectifying junction is not an ideal one, 
and a tiny reverse leakage current flows through the junction. This 
reverse leakage current is typically several order of magnitude greater 
than the thermionic flux within the spheres and allows the transfer of 
charges and potential drop to the terminal leads~\cite{dab2}.

If we short the terminal leads through a resistor $R$, then it should be 
possible to exploit the potential drop $V$, for example generating heat 
through $R$ (Joule effect). It is possible to show that the power output 
per unit area of the inner sphere ${\cal P}_s$ is given by

\begin{equation}
{\cal P}_s=\frac{2\pi eV_s}{c^2}
\biggl(\frac{kT}{h}\biggr)^3\Biggl(\overline{\eta}_{in}\int_{\frac{eV_s +
\phi_{in}}{kT}}^\infty \frac{x^2 dx}{e^{x}-1} - 4\overline{\eta}_{ext}
\int_{\frac{\phi_{ext}}{kT}}^\infty \frac{x^2 dx}{e^x-1}
\Biggr). 
\label{eq9}
\end{equation}

\begin{figure}[t]
\centerline{\includegraphics[width=10cm]{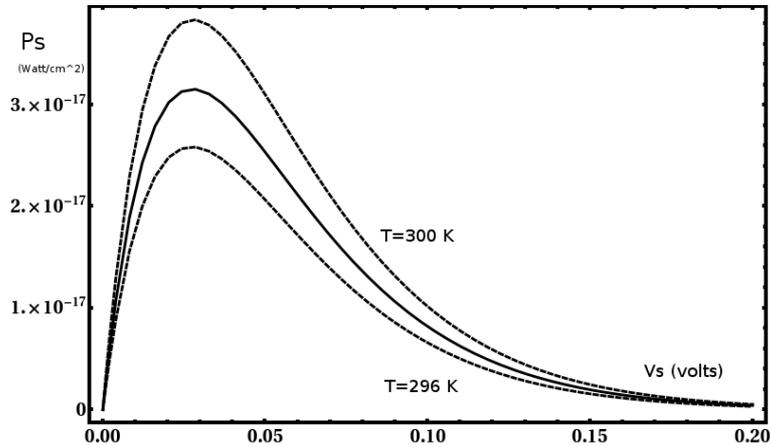}}
\caption{Power output per unit surface area of the inner sphere, 
${\cal P}_s$, against voltage drop $V_s$  across resistor $R$. 
Power outputs for $T=300\,$K and for $T=296\,$K are also shown.
Adapted from~\cite{dab2}.}
\label{fig6}
\end{figure}

$V_s$ is the steady-state potential drop across $R$ and across the capacitor 
after the current stabilizes. Fig.~\ref{fig6} shows a plot of the above 
function, in terms of the potential drop $V_s$.

For the capacitor described in Fig.~\ref{fig4} with $a=10\,$cm 
($b=20\,$cm), we have $P_{max}\approx 10^{-14}\,$Watts. This is a quite 
microscopic output but for laboratory devices, suitably designed for 
such tests, it should not be a concern, provided that an extremely 
sensitive, high input impedance electrometer (e.g.~Keithley 616, 617,
6514) is used as measurement equipment.

Moreover, some possible confounding factors, like thermocouple and 
Thomson effects, can be reduced or canceled out through a proper design 
of the laboratory devices~\cite{dab2}.

One may wonder why such an effect has not been observed before within 
vacuum tubes. After all, centimeter vacuum tubes have been widely used 
in electronic devices (phototubes, photomultipliers, radios, TVs, etc.) 
over a long time before the discovery of silicon (photo)diodes. 
A possible answer is that the thermo-charged capacitor is an ultra-high 
output impedance source (Tera-ohms or tens of Tera-ohms) and the effect 
described here is a really tiny one (power output $\approx 
10^{-14}\,$Watts), to the point that it may be easily masked by a voltage 
offset due to electrometer input bias current during direct 
measurements or, if detected, may be confused with other known thermionic 
effects (thermocouple/Thomson effects). Moreover, the commercial vacuum 
(photo)tubes are not properly designed to measure it. Laboratory tests of 
thermo-charged capacitors are currently under study.

\section{Concluding Remarks}

In this paper we have argued that we still do not have any fully cogent 
argument (known fundamental physical laws) to exclude macroscopic 
violation of the second law of thermodynamics in its classical 
formulation (Kelvin--Planck and Clausius postulates). Even Landauer's 
information-theoretic principle seems to fall short of the initial 
expectations of being the fundamental `physical' reason of all Maxwell's 
demons failure. We also described two experimental challenges which have 
been proposed in recent years and the physics behind them.

However, without unambiguous experimental results (which are currently 
lacking) it is difficult to say whether these experiments actually 
challenge or violate the thermodynamic second law, though the theory 
behind them appears to be sound and as yet unchallenged. Concerning the 
thermo-charged capacitor challenge, for `successful violation' we mean 
production of voltage/current ascribable only to the thermo-charging 
process beyond any reasonable doubt (and not to other spurious effects 
like Thomson/Seebeck effects or to measurement interferences).
 
The impact of any proven success would be understandably enormous from 
the point of view of basic principles. However, the present author does 
not believe that, at this stage, such successes would also have profound 
practical consequences. The power output is so minuscule that it is 
unthinkable to extract usable work from environmental heat.

Surely, they would shed new light on the possible distinction between 
thermodynamic entropy (classical thermodynamics) and statistical 
entropy. As hinted to in Section~2.2, statistical entropy remains 
unaffected by possible proven successes of these challenges. Even if it 
turns out that thermodynamic second law is violable, the breakage of a 
glass, the mixing of milk and coffee and the mixing of two distinct 
gases, for instance, will always be ``irreversible'' processes when 
taken as such (namely, not aided in some way by any of the thermodynamic 
second law violating devices described above). The direction of such 
processes always is in the sense of increasing Boltzmann's entropy. In 
case of future positive results, we are sure that this last embryonic 
thought about such a distinction will deserve further investigation.

\section*{Acknowledgements}

This work has been partially supported by the Italian Space Agency under 
ASI Contract N$^\circ$ 1/015/07/0. The author gratefully acknowledges 
Dr.ssa Assunta Tataranni for her advice and Prof. John D.~Norton 
(University of Pittsburgh) for his insightful suggestions. The precious 
comments and suggestions of two anonymous reviewers are also kindly 
acknowledged.

\end{document}